\documentclass[aps,twocolumn,reprint]{revtex4} % Revtex preprint format

\usepackage{graphicx} %% for loading postscript figures
\usepackage{amssymb,amsmath}
\usepackage{nomencl} % Build nomenclature automatically
  
\makenomenclature % For making nomenclature

\begin{document}

\title{Conduction in jammed systems of tetrahedra}

%%% first author
\author{Kyle C. Smith}
\email{kyle.c.smith@gmail.com}
\affiliation{Birck Nanotechnology Center and School of Mechanical Engineering,\\Purdue University, West Lafayette, Indiana 47907, USA}

%%% second author
\author{Timothy S. Fisher}
\email{tsfisher@purdue.edu}
\affiliation{Birck Nanotechnology Center and School of Mechanical Engineering,\\Purdue University, West Lafayette, Indiana 47907, USA}

%%%%%%%%%%%%%%%%%%%%%%%%%%%%%%%%%%%%%%%%%%%%%%%%%%%%%%%%%%%%%%%%%%%%%%
\begin{abstract}
{Control of transport processes in composite microstructures is critical to the development of high performance functional materials for a variety of energy storage applications.  The fundamental process of conduction and its control through the manipulation of granular composite attributes (e.g., grain shape) are the subject of this work.  We show that athermally jammed packings of tetrahedra with ultra-short range order exhibit fundamentally different pathways for conduction than those in dense sphere packings.  Highly resistive granular constrictions and few face-face contacts between grains result in short-range distortions from the mean temperature field.  As a consequence, `granular' or differential effective medium theory predicts the conductivity of this media within 10 \% at the jamming point; in contrast, strong enhancement of transport near interparticle contacts in packed-sphere composites results in conductivity divergence at the jamming onset.  The results are expected to be particularly relevant to the development of nanomaterials, where nanoparticle building blocks can exhibit a variety of faceted shapes.}
\end{abstract}

\maketitle

%%%%%%%%%%%%%%%%%%%%%%%%%%%%%%%%%%%%%%%%%%%%%%%%%%%%%%%%%%%%%%%%%%%%%%

\section{Introduction}

Heterogeneous materials represent an important class of practical feedstocks in a variety of applications related to energy technology.  Examples of such applications include waste heat recovery with thermoelectrics \cite{ShaARMR2011}, electrical energy storage with composite battery electrodes \cite{SmiPCCP2012}, and hydrogen storage with solid-state materials \cite{SclNat2001,SmiIJHE2012}.  Nanostructuring in particular is expected to enable rapid advances in the performance of materials for such applications \cite{BaxEES2009}.  Transport of charge and/or heat are essential to the successful performance of these materials, but a theoretical foundation for understanding the influence of microscopic constituent attributes and processing conditions on transport processes is lacking.

The investigations of Maxwell \cite{Max1873} and Lord Rayleigh \cite{RayPMag1892} pioneered the theoretical study of heterogeneous material properties, considering transport through systems of disks embedded in a host.  The effective conductivity of heterogeneous materials composed of smooth particles (most commonly spheres) have subsequently received extensive theoretical attention (see \cite{TorRHM2002}).  Most notably, effective medium approximations (EMA) of various types have been developed to model transport properties, including Maxwell-Garnett (MG) EMA \cite{Max1873}, self-consistent Bruggeman EMA \cite{BruAnn1935}, and granular (also referred to as differential or self-similar) EMA \cite{YonJAP1983,SenGeo1981}.  Modifications of the MG-EMA have even been made to incorporate thermal boundary resistance \cite{HasJCM1987,NanJAP1997}, but such extensions are primarily limited to low inclusion phase densities.  A differential EMA with thermal boundary resistance was developed in Ref. \cite{EveAct1992} to be applicable at high inclusion phase densities.  

Additionally, phenomenological correlations for the conductivity of packed beds have been developed based on simplified micromechanical contact models between particles (see \cite{TsoCEP1987}).  Despite these advances, little understanding has been gained regarding the effects of particle shape and microstructure on the transport properties of heterogeneous media.  Granular jamming processes, that are hereafter defined, provide physically-based means by which to generate microstructures.  The jamming threshold density $\phi_J$ in a granular medium represents the particular density $\phi$ \nomenclature[phi]{$\phi$}{density, -} (i.e., granular volume fraction) at which the medium transitions from an unjammed, fluid-like phase to a jammed phase that is solid but fragile \cite{CatPRL1998}.  Recently, a diversity of dense microstructures composed of tetrahedral building blocks has been discovered, including quasi-crystal and glassy structures through thermal excitation and densification \cite{AkbNat2009} and amorphous structures via athermal jamming \cite{SmiPRE2010,SmiPRE2011}.

Amorphous structures with short-range granular phase continuity \cite{SmiPRE2011} are particularly relevant to thermoelectric applications for which low thermal conductivity is desirable \cite{ShaARMR2011}, and also ball-mill processing of bulk thermoelectric materials that generates anisometric faceted nanoparticles has yielded high thermoelectric figure-of-merit through thermal conductivity reduction \cite{MaNL2008}.  Additionally, recent studies have shown that the thermoelectric power factor of a composite with dilutely embedded spherical nanoparticles is enhanced relative to a homogeneous impurity-doped material \cite{ZebNL2011}.  Ballistic-diffusive phonon transport studies have been conducted to assess the importance of particle ordering \cite{JenJHT2008}, particle shape \cite{WanJAP2011,HsiJAP2009}, and topological phase continuity \cite{ZhoPRB2010}.  As yet, no analysis of composites whose structure has physical origin, e.g., through sequential consolidation, has been reported.

In this work we explore the dependence of effective conductivity on granular density and grain conductivity (relative to pore conductivity) in jammed assemblies of tetrahedral grains with ultra-short range granular order \cite{SmiPRE2010,JaoPRL2010} and finite extent of the granular phase \cite{SmiPRE2011}.  While heat flow in dense packings of highly conducting spheres is localized in the vicinity of contacts \cite{KelJAP1963}, we find that jammed tetrahedra exhibit a temperature field with small deviations from the mean field.  Because of the insensitivity to the finite extent of face-face clusters, jammed tetrahedra exhibit bounded effective conductivity even for superconducting granular phases.  These results suggest that the paths through which heat flows in jammed faceted granular media are fundamentally different than those in jammed sphere systems.

\printnomenclature

\section{Methods}

\subsection{Jamming}

\begin{figure*}
\centering
\includegraphics[width=0.875\textwidth]{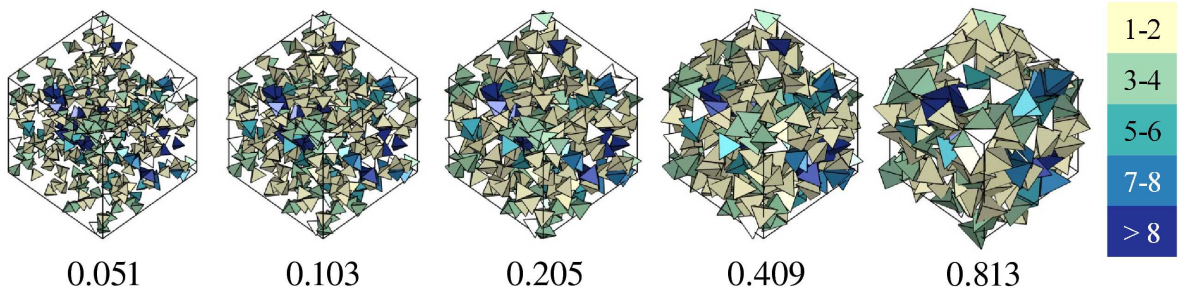}
\caption{Consolidation of $400$ tetrahedra.  Density is indicated below each image and color indicates the number of particles belonging to the face-face cluster of a given particle at the jamming threshold density of $\phi_J=0.634$ for this system.  Boundaries of the periodic supercell are indicated by black edges of the surrounding cube.}
\label{fig:consolidation}
\end{figure*}

Jammed systems having $25$, $100$, $400$, and $1600$ tetrahedral grains are considered in the present study.  The mechanics of these particles are modeled as pair-wise elastic interactions in which kinetic energy is neglected \cite{SmiPRE2010,SmiPRE2011}.  Jamming of soft mechanical systems comprised of $N$ such particles is simulated through controlled consolidative and expansive strain from initially dilute random states \cite{SmiPRE2010}.  Structural relaxation is performed at each density yielding mechanical stability of the system's internal degrees of freedom when jammed \cite{SmiPRE2010,SmiPRE2011}.  Sample states during the consolidation process are depicted for a system of $400$ tetrahedra in Fig.~\ref{fig:consolidation}; a series of fluid-like, low-density states precedes the mechanically stable jammed configuration at $\phi=0.813$.  The jamming threshold density of each system is estimated via asymptotic expansion toward the jamming point \cite{SmiPRE2010}, which is determined as $\phi_J=0.634$ for the $400$ particle system.  Sub-jamming point configurations were subsequently prepared by affine expansion and structural relaxation of jammed systems to facilitate continuum conduction simulation through the heterogeneous material.   Face-face clusters in systems of $25$ and $100$ tetrahedra were constrained to be in contact during expansion, while no such constraints were imposed for systems of $400$ and $1600$ tetrahedra.  

\subsection{Transport}

Conduction is simulated through the heterogeneous domain via numerical solution of the temperature field $T$ \nomenclature[T]{$T$}{temperature, K} governed by the steady heat diffusion equation:

\begin{equation}
 \nabla \cdot (\kappa \nabla T) = 0,
 \label{eq:heateq}
\end{equation} 

\noindent Heterogeneity manifests through the spatially dependent local conductivity $\kappa$, \nomenclature[ka]{$\kappa$}{local conductivity, W/m-K} represented by $\kappa_g$ \nomenclature[kapg]{$\kappa_g$}{grain conductivity, W/m-K} and $\kappa_p$ \nomenclature[kapg]{$\kappa_p$}{pore conductivity, W/m-K} in granular and pore phases, respectively.  In all cases, grain and effective conductivities are implicitly normalized by the pore-phase conductivity $\kappa_{p}$, such that the presented values are dimensionless.  Interfacial thermal boundary resistance (TBR) and interparticle contact resistances are neglected in the present work.  While TBR can be an important factor in overall heat conduction through heterogeneous media \cite{HasJCM1987,NanJAP1997}, the canonical EMA models to which we later compare the present results also exclude such effects.

At macroscopic scales the heterogeneous medium exhibits average heat flux $\langle \vec{q}'' \rangle$ \nomenclature[qpp]{$\vec{q}''$}{average heat flux vector, W/m$^2$} given by the generalized form of Fourier's law:

\begin{equation}
 \langle \vec{q}'' \rangle = -\boldsymbol{\kappa}(\nabla T)_h.
 \label{eq:kappatensor}
\end{equation}

\noindent where $(\nabla T)_h$ \nomenclature[DTh]{$(\nabla T)_h$}{homogenized temperature gradient vector, K/m} is the macroscopic, homogenized temperature gradient and $\boldsymbol{\kappa}$ is the effective conductivity tensor.  Periodic temperature fall boundary conditions are employed to probe $\boldsymbol{\kappa}$:

\begin{equation}
  T(\boldsymbol{r})=T(\boldsymbol{r}+\boldsymbol{r}_0)+ (\nabla T)_h \cdot \boldsymbol{r}_0, 
\end{equation}

\noindent where $\boldsymbol{r}$ \nomenclature[r]{$\boldsymbol{r}$}{position vector, m} is position in the heterogeneous medium and $\boldsymbol{r}_0$ \nomenclature[r0]{$\boldsymbol{r}_0$}{integer supercell vector combination, m} is an integer combination of principal lattice vectors defining the periodic supercell.  All systems in the present work were jammed with a cube-shaped supercell, as in Ref.~\cite{SmiPRE2010}, and the same supercell lattice vectors define the periodicity in transport simulations.

The conductivity tensor $\boldsymbol{\kappa}$ describes the response of granular on macroscopic scales.  To determine this quantity three unique boundary value problems ($1$, $2$, and $3$) with linearly independent homogenized temperature gradients [$(\nabla T)_{h1}$, $(\nabla T)_{h2}$, and $(\nabla T)_{h3}$] are imposed on each distinct microstructure (i.e., a system at a given density) to determine the resulting average heat flux vectors [$\langle \vec{q}'' \rangle_1$, $\langle \vec{q}'' \rangle_2$, and $\langle \vec{q}'' \rangle_3$].  By manipulating Eq.~\ref{eq:kappatensor} one can show that the effective conductivity tensor $\boldsymbol{\kappa}$ is fully specified from the data of these three boundary value problems:

\begin{equation}
 \boldsymbol{\kappa}  = -\boldsymbol{Q} \boldsymbol{G}^{-1},
\end{equation}

\noindent where each column in the matrix $\boldsymbol{Q}$ \nomenclature[Q]{$\boldsymbol{Q}$}{heat flux matrix, W/m$^2$} is the heat flux vector for a given boundary value problem (i.e., $\boldsymbol{Q}=\left[\langle \vec{q}'' \rangle_1 \langle \vec{q}'' \rangle_2 \langle \vec{q}'' \rangle_3\right]$) and the same column of matrix $\boldsymbol{G}$ \nomenclature[G]{$\boldsymbol{G}$}{gradient matrix, K/m} is the corresponding homogenized temperature gradient of that boundary value problem (i.e., $\boldsymbol{G}=[\left(\nabla T)_{h1} (\nabla T)_{h2} (\nabla T)_{h3}\right]$).  An orthonormal set of homogenized gradients along Cartesian axes is employed presently, in which case the gradient matrix $\boldsymbol{G}$ is the $3 \times 3$ identity matrix.

\begin{figure}
\centering
\includegraphics[width=0.23\textwidth]{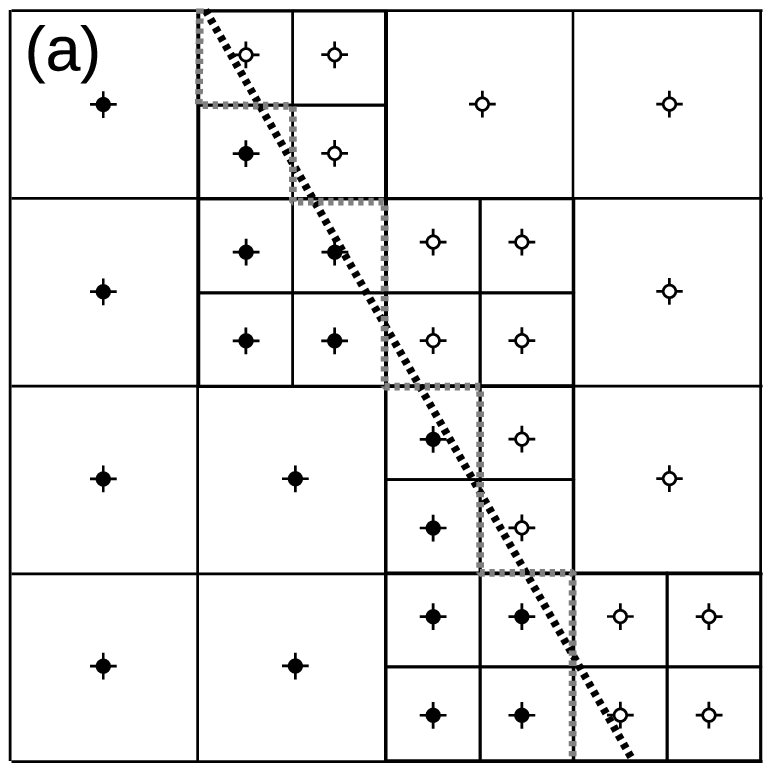}
\includegraphics[width=0.22\textwidth]{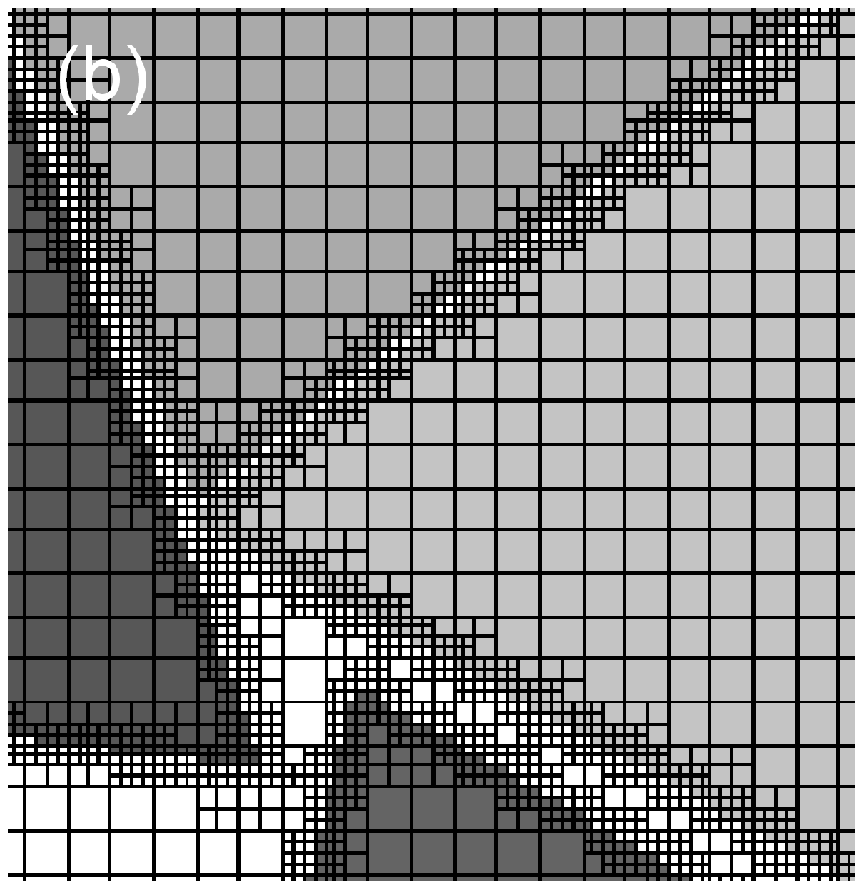}
\caption{Reconstructed interface geometry.  (a) Phase of a given cell is determined by that of its centroid and is indicated by the centroid's color.  The true and reconstructed interfaces are depicted in black and gray, respectively.  (b) Sample cross-section of the three-level refined mesh for a system of $400$ tetrahedra at $\phi=0.611$.  The pore phase is indicated in white, while distinct grains are shaded.}
\label{fig:octschematic}
\end{figure}

For transport simulations, grain-pore interfaces were represented as stair-stepped boundaries, as depicted in Fig.~\ref{fig:octschematic}(a), from which discrete equations were developed utilizing the finite volume method.  To utilize computer memory efficiently while achieving adequate grain-pore interface resolution, octree-based refinement was employed.  Cells in the finest octree level were determined by finding the intersection of cell edges and faces with clusters of tetrahedra.  A sample cross-section of the three-level octree mesh for $400$ nearly jammed tetrahedra is depicted in Fig.~\ref{fig:octschematic}(b).  Refinement was implemented through a cell neighbor detection scheme aided by three-dimensional sparse data structures \cite{ndSparse}.

\begin{figure}
\centering
\includegraphics[width=0.4\textwidth]{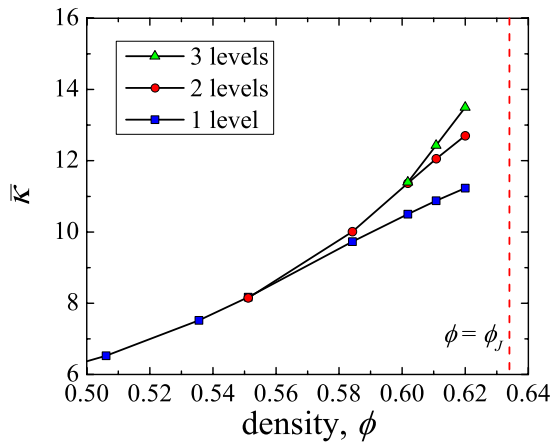}
\caption{Octree refinement of $\bar{\kappa}$ as a function of density for $400$ tetrahedra with $\kappa_g=10^3$.  For each data point the coarsest level of the mesh has a resolution of $\Delta x = 0.025 (N V_g/\phi)^{1/3}$, where $\Delta x$ is the length of coarse cube-shaped finite volume cells, $N$ is the number of particles, and $V_g$ is the volume of an individual grain.  The position of the jamming point is marked by the red dashed line.}
\label{fig:refstudy}
\end{figure}

In general, increased resolution is required to resolve pore phase gaps between grains approaching face-face contact at densities approaching the jamming point ($\phi \rightarrow \phi_{J}^{-}$).  To prevent artificial granular phase continuity, such cells are considered as pore phase.  This scheme yields topologically consistent reconstructed interfaces in the limit of infinite refinement.  The errors resulting from approximate meshes with finite refinement dominate over discretization errors.  To verify the adequacy of interfacial resolution at a given density $\phi$, several levels of octree refinement were considered (Fig.~\ref{fig:refstudy}).  Systems with highly conducting grains ($\kappa_g=10^3$) were chosen for this purpose because of their high sensitivity to refinement.  As evidenced by Fig.~\ref{fig:refstudy}, the conductivity with one octree level (i.e., a uniform Cartesian mesh) is satisfactory at $\phi=0.550$, while two octree levels yield adequate convergence at $\phi=0.600$.  Such studies were performed for various microstructural configurations to ensure adequate convergence; in all cases the coarsest level mesh had a size of $\Delta x = 0.025 (N V_g/\phi)^{1/3}$, where $V_g$ is the volume of an individual grain.  We present results for refined meshes having mean conductivity that differs by less than 3\% with that obtained with one level greater refinement.  Such explicit refinement studies were performed only on systems of 25, 100, and 400 tetrahedra, and the results were employed to mesh the remaining systems suitably.  As a result of increased resolution requirements near the jamming and physical computer memory limitations, the conductivities of small systems (25 and 100 tetrahedra) were simulated within 1\% of the jamming threshold density, while larger systems were simulated farther from the jamming point.

\begin{figure}
\centering
\includegraphics[width=0.25\textwidth]{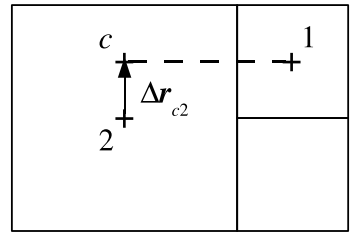}
\caption{Stencil employed for discretization of flux at faces between cells of dissimilar size.}
\label{fig:discells}
\end{figure}

The octree-based refinement technique greatly simplifies computational access to grid topology, but lack of regular structure in the mesh complicates the determination of fluxes between adjacent cells from the standard interpolation formulas (see Ref.~\cite{JalCHT2003}).  Here, a simplified discretization scheme is introduced to interpolate flux between cells of dissimilar size to circumvent this issue; for adjacent cells of equal size the scheme reduces to the central difference discretization.  An example of the dissimilar case is shown in Fig.~\ref{fig:discells}, where cell $2$ is larger than cell $1$.  Flux at the boundary between the two cells is determined assuming temperature at a ghost point $c$ is known.  Because point $c$ is eccentric to the centroid of cell $2$ and a cell-centered scheme is sought, the value of temperature at $c$ must be expressed in terms of those at the centroid of cell $2$ and other neighboring cells (e.g., in Ref. \cite{JohaJCP1998}).  The aggregation-based algebraic multigrid (AGMG) method was employed to solve the discrete set of equations in the present work \cite{NotayETN2010,NotayTech2010,NotayTech2011}, and the algorithm therein requires diagonal dominance of the corresponding system matrix.  A discretization explicitly including the temperature at point $c$ violates the Scarborough criterion and does not converge.  Therefore, an iterative approach to express the temperature $T_c$ at $c$ is employed as:

\begin{equation} 
T_c = T_2+ (\nabla T^*)_2 \cdot \Delta \boldsymbol{r}_{c2},
\label{eq:corr}
\end{equation}

\noindent where $T_2$ is the temperature at the centroid of cell $2$ and $\Delta\boldsymbol{r}_{c2}$ \nomenclature[Drc2]{$\Delta \boldsymbol{r}_{c2}$}{correction position vector, m} is the position vector extending from centroid $2$ to point $c$.  $(\nabla T^*)_2$ is the gradient of cell $2$ reconstructed from the heat fluxes computed from the most recent global solution obtained with AGMG.  Eq.~\ref{eq:corr} is employed to correct the flux between cells $1$ and $2$, expressed via linear interpolation, and is lumped into the residual vector of the linear system comprising the finite volume discretization of Eq.~\ref{eq:heateq}.  Typically five iterations of this global correction procedure are required to converge to a relative conservation residual of less than $10^{-5}$.

\section{Results}

\begin{figure}
\centering
\includegraphics[width=0.5\textwidth]{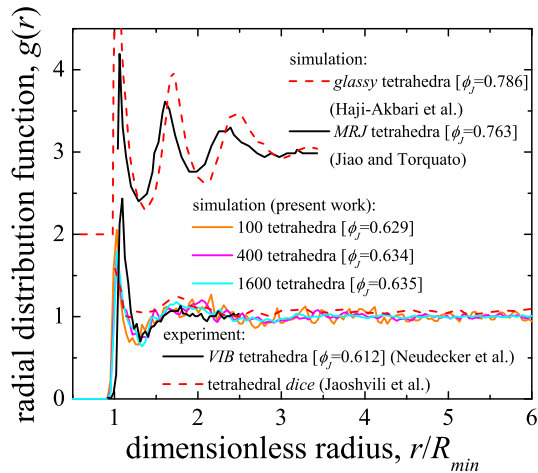}
\caption{Radial  distribution function of the present systems for various system sizes.  Experimental \cite{NeuAR2012,JaoPRL2010} and simulated \cite{AkbNat2009,JiaPRE2011} values of other contemporary works are shown as well.  The system of $25$ tetrahedra is excluded because of insufficient sampling.  In brackets are listed the jamming threshold densities.  $R_{min}$ \nomenclature[Rmin]{$R_{min}$}{minimum centroidal separation, m} is the minimum possible centroidal separation between contacting tetrahedra.}
\label{fig:RDF}
\end{figure}

For systems of 25, 100, 400, and 1600 particles we estimate the jamming threshold density $\phi_J$ to be $0.628$, $0.629$, $0.634$, and $0.635$, respectively; coincidentally these values are remarkably close to the jamming density of athermal soft spheres, $\sim 0.64$ \cite{HerPRE2003}.  Each of the present systems exhibits translational order very similar to that observed experimentally for dense packings of tetrahedra \cite{NeuAR2012} and tetrahedral dice \cite{JaoPRL2010}, validating the present jamming model.  In contrast, the simulated values of other works via Monte Carlo based methods \cite{AkbNat2009,JiaPRE2011} exhibit much higher jamming threshold densities and different structures than those of the present work.  The radial distribution function of the present systems converges with system size, as displayed in Fig.~\ref{fig:RDF}.  This observation suggests that the translational arrangement of 100 jammed tetrahedra and larger jammed systems are very similar.

\begin{figure}
\centering
\includegraphics[width=0.5\textwidth]{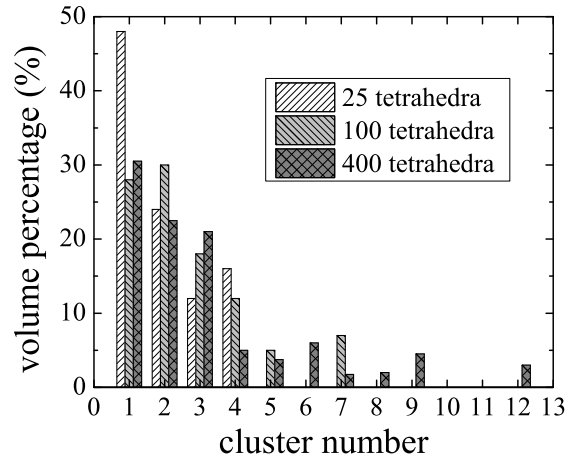}
\caption{Cluster number distribution for jammed systems of tetrahedra of varying system size.}
\label{fig:clsnum}
\end{figure}

In the absence of contact resistance resulting from roughness between face-to-face contacting particles, clusters formed through networks of face-to-face contacting particles transmit heat as a continuous solid phase.  Therefore, cluster morphology is expected to affect the macroscopic transmission of heat through the heterogeneous medium.  In the $400$ tetrahedra system the variety of jammed cluster sizes and shapes is clearly evident (Fig.~\ref{fig:consolidation}).  The fraction of granular volume represented by clusters of a given number of grains (i.e., cluster number) is presented in Fig.~\ref{fig:clsnum} for each jammed system.  From this distribution the finite size of the $25$ tetrahedra system is apparent, as it exhibits the largest fraction of monomers (i.e., non-clustered particles).  

Systems of $100$ and $400$ tetrahedra exhibit very similar cluster number distributions except for deviations at large cluster sizes; in both systems monomer, dimer, and trimer clusters represent nearly 80 \% of the granular volume.  The similarities in cluster topologies are consistent with the observations of similar radial distribution functions among systems with as few as $100$ tetrahedra.  Despite dissimilarity among large clusters, the maximum cluster extent relative to system size decreases as system size increases \cite{SmiPRE2011}.  Primarily this phenomenon results from the various jammed cluster morphologies in the highest density configuration depicted in Fig.~\ref{fig:consolidation}, only the longest of which exhibit slender, chain-like morphologies \cite{SmiPRE2011}.  None of the highly-ordered cluster morphologies (e.g., tetrahedral helices and pentagonal dipyramids) predicted to form in tetrahedral crystals, quasi-crystals, and glasses (see \cite{AkbNat2009}) appear in the present amorphous jammed systems.  In all systems the mean cluster number is small ($\sim2$), and therefore, similar behavior among these systems' mean conductivities is expected. 

Periodic boundaries consistent with those in the jamming simulation were employed to determine the effective thermal (and by analogy, electrical) conductivity tensor $\boldsymbol\kappa$.  Though systems of tetrahedra are expected to be microstructurally isotropic in the infinite system limit, the finite systems under study here exhibit anisotropy due to the orientation of a small, finite number of clusters in each system.  Anisotropy of $\boldsymbol\kappa$ in finite systems is therefore reflective of uncertainty in multiple random realizations of conduction processes in such media.  The components of conductivity $\kappa_{1}$, $\kappa_{2}$, and $\kappa_{3}$ (in descending order of magnitude) along the principal directions ($1$, $2$, and $3$, respectively) \nomenclature[kap123]{$\kappa_{1,2,3}$}{principal conductivities, W/m-K} are determined by the eigen decomposition of $\boldsymbol\kappa$.  Two independent measures of conduction in the effective medium are considered -- mean conductivity $\bar{\kappa}$ \nomenclature[kapb]{$\bar{\kappa}$}{mean conductivity, W/m-K} and deviatoric conductivity $\Delta \kappa$, \nomenclature[Dkap]{$\Delta \kappa$}{deviatoric conductivity, W/m-K} defined as:

\begin{equation}
  \bar{\kappa} = (\kappa_1+\kappa_2+\kappa_3)/3,
\end{equation}

\begin{equation}
  \Delta \kappa = \kappa_{1} - \kappa_{3}.
\end{equation}

\begin{figure}
\centering
\includegraphics[height=0.4\textwidth]{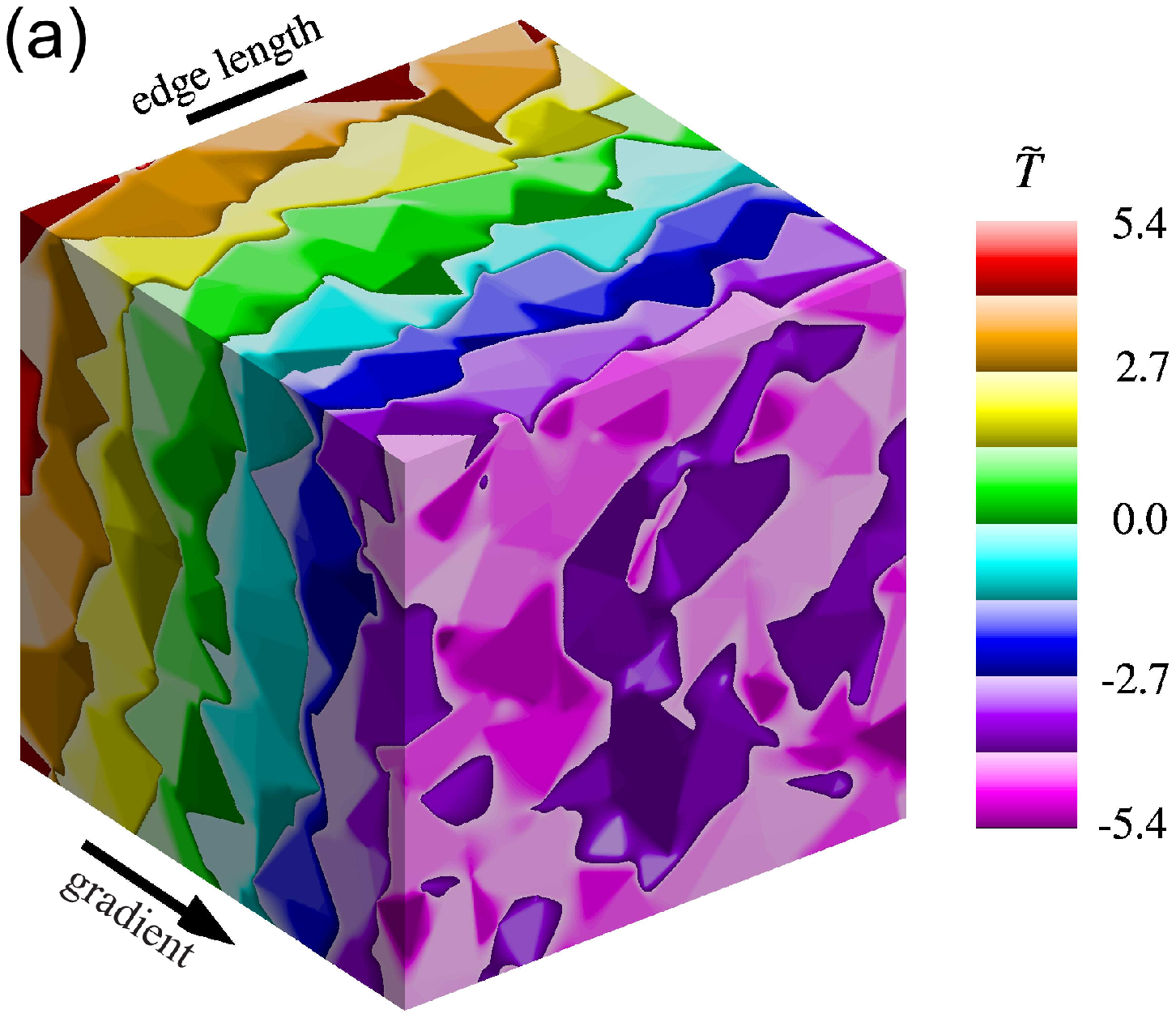}\\
\includegraphics[width=0.5\textwidth]{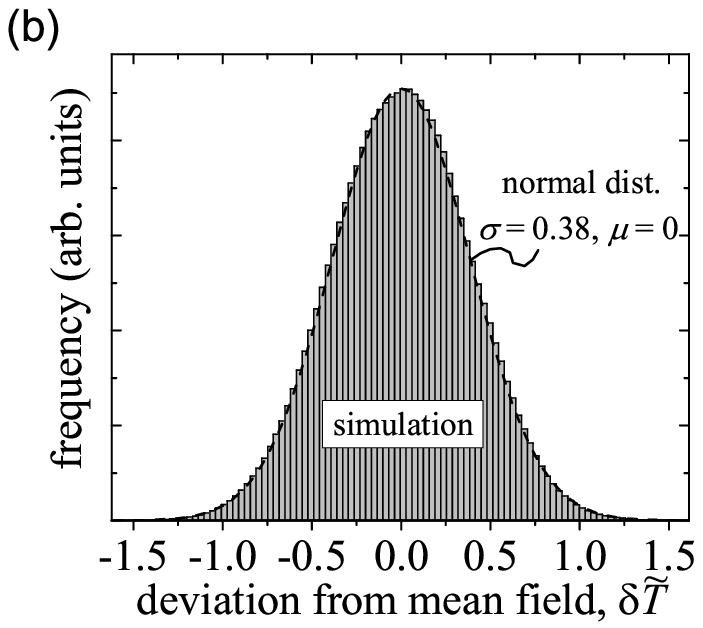}
\caption{(a) Normalized temperature field $\tilde{T}$ \nomenclature[Tt]{$\tilde{T}$}{normalized temperature field, -} for 400 tetrahedra with $\kappa_g=10^3$ at $\phi=0.611$.  The field is induced by a homogenized temperature gradient $(\nabla T)_h$ applied to the system along the indicated direction.  The length scale of a tetrahedron edge is indicated as a reference for the size of an individual grain. (b) Histogram of the deviation $\delta\tilde{T}$ \nomenclature[dT]{$\delta\tilde{T}$}{normalized deviation from mean temperature field, -} from the mean normalized temperature field.  The fitted normal distribution is shown along with its standard deviation $\sigma$ \nomenclature[sig]{$\sigma$}{standard deviation, -} and mean value $\mu$ \nomenclature[mu]{$\mu$}{mean value, -}.}
\label{fig:tfieldtet}
\end{figure}
 
The temperature field $\tilde{T}$ for $400$ nearly superconducting tetrahedra is displayed in Fig.~\ref{fig:tfieldtet}(a), where $\tilde{T}$ is normalized by the characteristic temperature drop across an individual grain $|(\nabla T)_h|(V_g)^{1/3}$ and $V_g$ \nomenclature[Vg]{$V_g$}{individual grain volume, m$^3$} is the volume of an individual grain.  Fig.~\ref{fig:tfieldtet} illustrates that distortions from the mean temperature field have length scales on the order of tetrahedron edge lengths.  Thus, heterogeneous fluctuations in the temperature field of the largest system are small relative to the overall length scale of the system -- a direct consequence of the short range of face-face clusters [see Fig.~\ref{fig:clsnum}].  The short-range heterogeneous fluctuations are further evidenced by the distribution of deviations from the mean field $\delta\tilde{T}$ in Fig.~\ref{fig:tfieldtet}(b).  The normal distribution of field fluctuations is likely linked to the lack of translational order of grains reflected in the radial distribution function (Fig.~\ref{fig:RDF}).  The magnitude of this deviation is proportional to the spatial extent $\zeta$ \nomenclature[zeta]{$\zeta$}{cluster spatial extent, m} of polarized granular islands within the microstructure (i.e., $\zeta \sim \delta\tilde{T}$).  The distribution's short-range bounds (i.e., $|\delta\tilde{T}|<2$) suggest that the spatial extent of deviations is much smaller than the periodic system's length $L$, \nomenclature[L]{$L$}{periodic supercell length, m} because $|\delta\tilde{T}| < L/V_g^{1/3}=8.7$.  This behavior strongly contrasts with the field for systems with percolating clusters at jamming, where maximal distortions to the mean field are of the order of system size.

\begin{figure}
\centering
\includegraphics[width=0.5\textwidth]{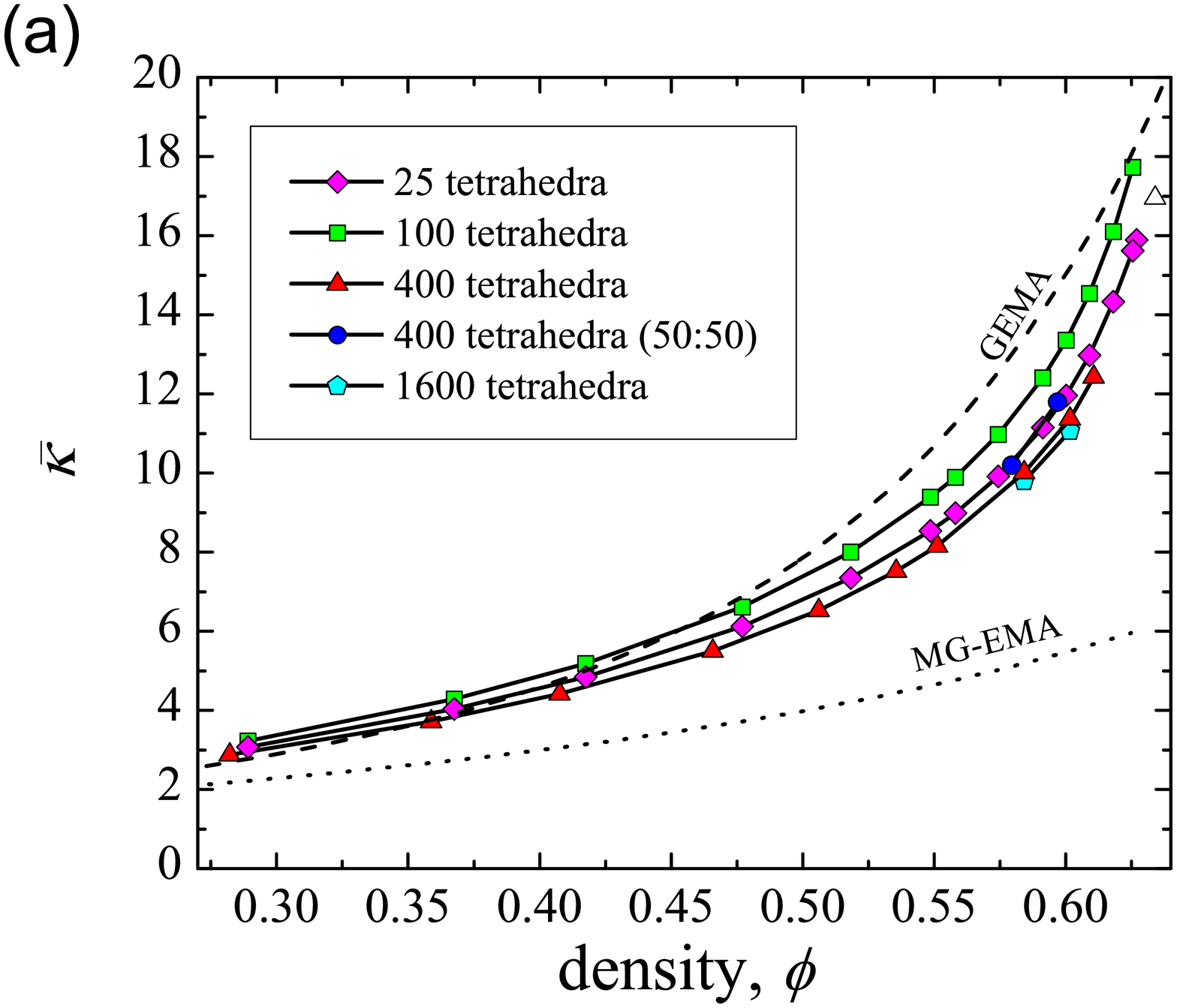}\\
\includegraphics[width=0.5\textwidth]{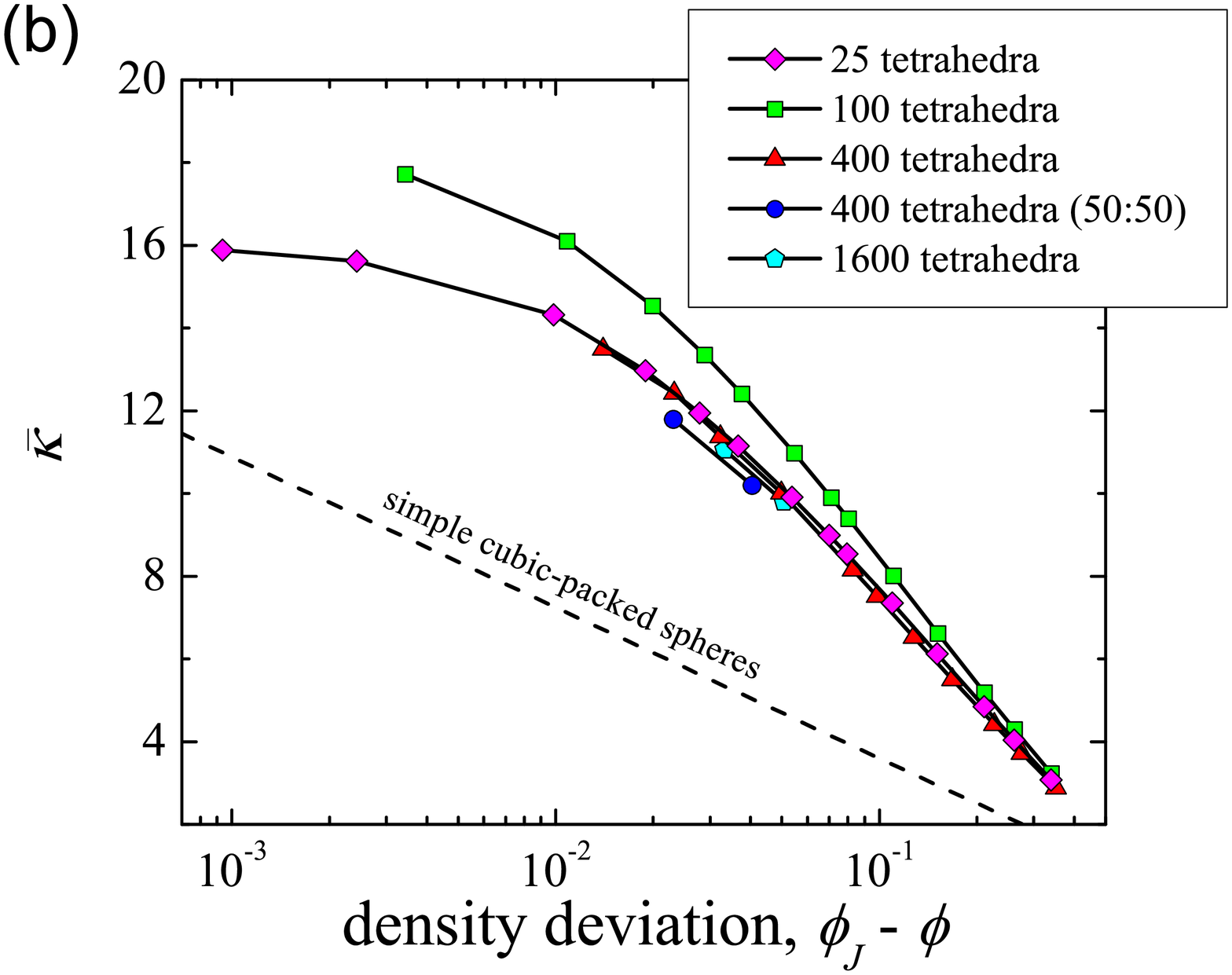}
\caption{Mean effective conductivity for tetrahedra with $\kappa_g=10^3$ as a function of (a) density and (b) density deviation, $\phi_J-\phi$, where $\phi_J$ is the jamming density for the respective system.  The open triangular symbol represents a cubic  polynomial extrapolation to the jamming point.  50:50 denotes the bidisperse mixture of tetrahedra described in the text.  For simple cubic-packed spheres $\phi_J=\pi/6$, the density at which spheres just touch.  GEMA refers to the granular effective medium approximation presented in \cite{YonJAP1983,SenGeo1981}, and MG-EMA refers to the Maxwell-Garnett effective medium approximation (see \cite{ChoEff1999}).  Data for simple cubic-packed spheres is from \cite{McPRSA1983} at low density and \cite{KelJAP1963} at high density.}
\label{fig:kphikDphi}
\end{figure}

Additionally, the tetrahedron and sphere exhibit the lowest and highest rotational symmetry, respectively, among regular convex shapes.  Thus, these shapes represent extremes in the spectrum of granular building blocks for heterogeneous materials.  A peculiar feature of dense sphere packings is the emergence of a superconducting composite when spheres begin to touch, which has been readily explored for lattice sphere packings \cite{McPRSA1983}.  This behavior is expected to be relatively insensitive to long-range granular ordering, because heat flow is localized in regions near the contacts between particles \cite{KelJAP1963}.  In contrast, unless tetrahedra contact via opposing faces, intersections between them at the jamming point have low-order topology (e.g., vertex or edge) that are expected to be highly resistive.

Though numerical limitations prevent the jamming point from being reached exactly, the study of systems that approach this point are expected yield understanding regarding the mean conductivity at that point.  To probe the effect of the jamming condition on conduction, configurations were studied with densities $\phi$ less than the jamming threshold density $\phi_J$.  Fig.~\ref{fig:kphikDphi}(a) shows the variation of mean effective conductivity $\bar{\kappa}$ with density $\phi$ in addition to the predictions of theoretical EMA models.  Despite the variations in system size, $\bar{\kappa}$ is very similar among all systems at a given density.  Though the so-called granular (or differential) EMA \cite{YonJAP1983,SenGeo1981} contains no adjustable parameters, $\bar{\kappa}$ of tetrahedral systems deviates from its prediction by at most 10 \% at the jamming point for $400$ tetrahedra; in contrast the MG-EMA does not follow the same trend.  Also, the lack of sub-critical scaling of $\bar{\kappa}$ with respect to density deviations from the jamming point confirms that conduction in systems of tetrahedra is relatively unaffected by the detailed structure of the grain phase [Fig.~\ref{fig:kphikDphi}(b)].  Tetrahedral systems contrast strongly with systems of spheres that exhibit logarithmic variation with respect to the density at which spheres touch (cf. \cite{KelJAP1963}).

\begin{figure}
\centering
\includegraphics[width=0.5\textwidth]{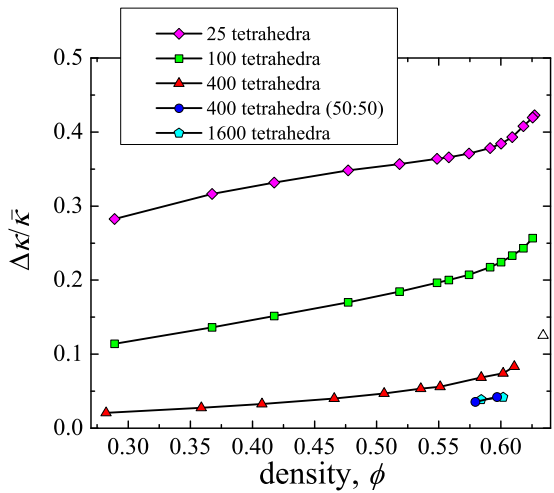}
\caption{Deviatoric conductivity $\Delta \kappa$ as a function of $\phi$ for $\kappa_g=10^3$.  The open triangular symbol represents a cubic polynomial extrapolation to the jamming point, and 50:50 denotes the bidisperse mixture of tetrahedra described in the text.}
\label{fig:devcond}
\end{figure}

\begin{figure}
\centering
\includegraphics[width=0.5\textwidth]{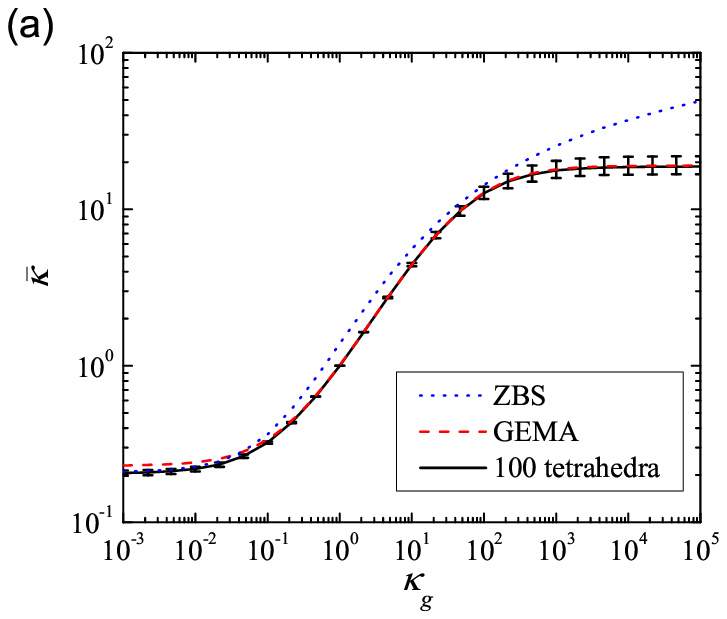}\\
\includegraphics[width=0.5\textwidth]{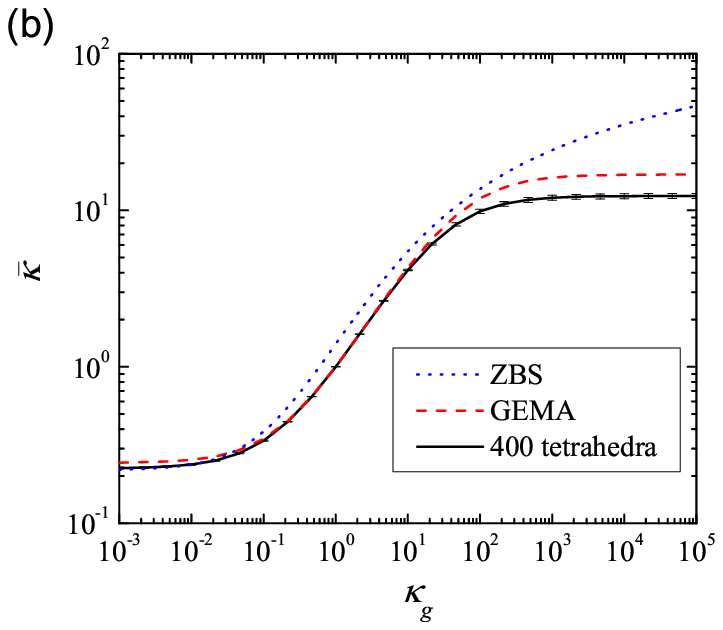}
\caption{Mean effective conductivity as a function of grain conductivity $\kappa_g$ for (a) $100$ tetrahedra at $\phi = 0.626$ and (b) $400$ tetrahedra at $\phi = 0.611$.  Error bars represent $\kappa_{1}$ and $\kappa_{3}$.  Theoretical data is presented based on the granular effective medium approximation (GEMA) in \cite{YonJAP1983,SenGeo1981} and the Zehner-Bauer-Schl\"under (ZBS) model (see \cite{TsoCEP1987}).}
\label{fig:kks_compare}
\end{figure}

The highly disordered structure and short range of clusters in systems of tetrahedra results in macroscopic response similar to that of granular effective media.  This effect is not unique to systems of monodisperse tetrahedra only, but is reflective of a wide class of granular systems composed of highly anisometric faceted grains.  To illustrate this point a mixture of $200$ tetrahedra with $V_g^{1/3}=1$ and $200$ tetrahedra with $V_g^{1/3}=1.4$ was considered; similar size ratios of spheres can function to frustrate structural effects that are peculiar to monodisperse systems (see \cite{HerPRE2003}).  This mixed system exhibited a jamming threshold density of $0.620$ and very similar effective thermal conductivity to that of monodisperse tetrahedra [see Figs.~\ref{fig:kphikDphi}(a,b)].  Fig.~\ref{fig:devcond} confirms that the primary effect of structural frustration is to reduce conduction anisotropy relative to monodisperse systems of the same size.  The shape of faceted particles need not be exactly tetrahedral to exhibit similar behavior to regular, monodisperse tetrahedra, as we have recently observed similar behavior in packings of irregular, polydisperse metal hydride particles \cite{SmiIJHE2012}.  In this light, conduction through systems of spheres is fundamentally different as a result of dominant local contact interactions. 

The deviatoric conductivity $\Delta \kappa$, which reflects conduction anisotropy, is displayed in Fig.~\ref{fig:devcond} for systems with nearly superconducting grains.  $\Delta \kappa$ depends strongly on system size -- a result of the small number of large clusters in small systems.  Thus, small systems conduct preferentially along such clusters, while in large systems, where many clusters of various orientations form, conduction is more isotropic.  This finite size effect is relatively absent in the $1600$ tetrahedra system and is expected to vanish as system size is further increased.  

Until this point granular systems of nearly superconducting tetrahedra have been considered, revealing close quantitative agreement with granular EMA predictions.  Fig.~\ref{fig:kks_compare} displays the variation of mean conductivity with grain conductivity $\kappa_g$.  The curve exhibits a sigmoidal shape resulting from insulating ($\kappa_g \rightarrow 0$) and superconducting ($\kappa_g \rightarrow \infty$) grain conductivity regimes.  In these respective limits the granular EMA deviates most from the effective conductivity of the tetrahedral system, while in the region of moderate grain conductivity ($0.1 < \kappa_s < 10$) the granular EMA agrees well with the present numerical results.  The decreased conductivity of jammed tetrahedra relative to the granular EMA prediction in the extreme limits of $\kappa_g$ is likely a result of the localized constriction of heat between contacting tetrahedra and not the difference in the polarizability of face-face clusters present in jammed packings and isolated spheres considered in the granular EMA.  In contrast, the popular packed bed conductivity correlation of Zehner, Bauer, and Schl\"under \cite{TsoCEP1987} fails to reproduce finite conductivity in the superconducting granular limit, as well as in the moderate grain conductivity regime.  

\section{Conclusions}

The agreement of the granular EMA with the present results is not merely a fortuitous coincidence.  EMA theories interpret heterogeneous materials as a product of phase-embedding processes.  The process implied by the granular EMA is asymmetric with respect to the embedding of pore and grain phases into model microstructures -- pores have perfect continuity, while grains are disjointed, at all densities \cite{YonJAP1983,SenGeo1981}.  It is not surprising, then, that the granular EMA, which has also shown agreement with experimental measurements of the electrical conductivity of sedimentary rock \cite{SenGeo1981}, models the conductivity of jammed tetrahedra well.  Robustness of this behavior to grain shape and size perturbations from monodisperse systems of tetrahedra, suggests that a wider class of jammed faceted granular media exists with similar structure and properties to tetrahedra.  The results also suggest that a granular EMA theory may be appropriate to model thermal boundary resistance effects in jammed systems of tetrahedra (e.g., in ref. \cite{EveAct1992}).  These findings also suggest that granular variants of other effective medium theories (e.g., the coherent potential approximation for electronic density of states in heterogeneous media) may be applicable to jammed tetrahedra and to the wider class of jammed faceted grains to which tetrahedra belong.  

\section{Acknowledgments}

K.C.S. thanks the U.S. National Science Foundation and the Purdue Graduate School for financial support.  Both authors thank the U.S. National Science Foundation's Office of International Science and Engineering for travel support that enabled illuminating foundational interactions on granular mechanics with Prof. Meheboob Alam of the J. Nehru Centre for Advanced Scientific Research.  Finally, the authors thank Prof. Jayathi Murthy, the PRISM center staff, and Phil Cheeseman of the Rosen Center for Advanced Computing for access to and support of computing resources.

% Here's where you specify the bibliography database file.
% The full file name of the bibliography database for this
% article is asme2e.bib. The name for your database is up
% to you.
\bibliography{../../mainbib.bib}

\end{document}